\documentclass[prl,twocolumn,showpacs,amssymb,superscriptaddress]{revtex4}
\usepackage{graphicx}

\begin{document}
\title{Hot electron driven enhancement of spin-lattice coupling in 4f ferromagnets observed by femtosecond x-ray magnetic circular dichroism}

\author{Marko Wietstruk}
\affiliation{Helmholtz-Zentrum Berlin f\"ur Materialien und
Energie GmbH, BESSY II, Albert-Einstein-Str. 15, 12489 Berlin,
Germany} \affiliation{Max-Born-Institut, Max-Born-Str.~2A, 12489
Berlin, Germany}

\author{Alexey Melnikov}
\affiliation{Freie Universit\"{a}t Berlin, Fachbereich Physik,
Arnimallee 14, 14195 Berlin, Germany}

\author{Christian Stamm}
\affiliation{Helmholtz-Zentrum Berlin f\"ur Materialien und
Energie GmbH, BESSY II, Albert-Einstein-Str. 15, 12489 Berlin,
Germany}

\author{Torsten Kachel}
\affiliation{Helmholtz-Zentrum Berlin f\"ur Materialien und
Energie GmbH, BESSY II, Albert-Einstein-Str. 15, 12489 Berlin,
Germany}

\author{Niko Pontius}
\affiliation{Helmholtz-Zentrum Berlin f\"ur Materialien und
Energie GmbH, BESSY II, Albert-Einstein-Str. 15, 12489 Berlin,
Germany}

\author{Muhammad Sultan}
\affiliation{Freie Universit\"{a}t Berlin, Fachbereich Physik,
Arnimallee 14, 14195 Berlin, Germany}

\author{Cornelius Gahl}
\affiliation{Max-Born-Institut, Max-Born-Str.~2A, 12489 Berlin,
Germany}

\author{Martin Weinelt}
\email[Corresponding authors: ] {weinelt@mbi-berlin.de}
\affiliation{Max-Born-Institut, Max-Born-Str.~2A, 12489 Berlin,
Germany}\affiliation{Freie Universit\"{a}t Berlin, Fachbereich
Physik, Arnimallee 14, 14195 Berlin, Germany}

\author{Hermann~A. D\"urr}
\affiliation{Helmholtz-Zentrum Berlin f\"ur Materialien und
Energie GmbH, BESSY II, Albert-Einstein-Str. 15, 12489 Berlin,
Germany} \affiliation{ PULSE Institute and Stanford Institute for
Materials and Energy Sciences, SLAC National Accelerator
Laboratory, Menlo Park, CA 94025, USA}

\author{Uwe Bovensiepen}
\email[] {uwe.bovensiepen@uni-due.de} \affiliation{Freie
Universit\"{a}t Berlin, Fachbereich Physik, Arnimallee 14, 14195
Berlin, Germany} \affiliation{Universit\"at Duisburg-Essen,
Fakult\"at f\"ur Physik, Lotharstr.~1, 47048 Duisburg, Germany}

\date{\today}

\begin{abstract}
Femtosecond x-ray magnetic circular dichroism was used to study
the time-dependent magnetic moment of $4f$ electrons in the
ferromagnets Gd and Tb, which are known for their different
spin-lattice coupling. We observe a two-step demagnetization with
an ultrafast demagnetization time of $750$~fs identical for both
systems and slower times which differ sizeably with $40$~ps for Gd
and $8$~ps for Tb. We conclude that spin-lattice coupling in the
electronically excited state is enhanced up to orders of magnitude
compared to equilibrium.
\end{abstract}

\pacs{78.47.J-, 71.38.-k, 75.70.Ak, 78.70.Dm}

\maketitle


Laser-induced magnetization dynamics has high potential for
ultrafast data-storage applications \cite{stanciu_PRL07} and a
microscopic understanding of the underlying processes is essential
for device optimization and tuning. In this context switching the
magnetic order by intense, ultrashort laser pulses explores the
speed limit of magnetic recording. Next to its technological
relevance magnetization dynamics driven by femtosecond (fs) laser
pulses challenges our microscopic understanding of magnetism:
(i) Bigot et al. \cite{bigot_NatPhys09} and Zhang et al.
\cite{zhang_NatPhys09} suggest that the light field is involved in
magnetization dynamics. (ii) Battiato et al. propose
superdiffusive spin transport as a mechanism of ultrafast
demagnetization \cite{battiato_PRL10}. (iii) Koopmans and
coworkers have developed an empirical model based on spin-orbit
mediated electron spin-flip scattering. Their concept
implies a material dependent demagnetization time and connects
itinerant and rare earth ferromagnets \cite{koopmans_NatMat10}.
Ultrafast laser-induced magnetization dynamics has been
established for the $3d$ metals and a number of alloys
\cite{ju_PRL04,thiele_APL04,
ogasawara_PRL05,bartelt_APL2007,walowski_PRL08,
malinowski_NatPhys08,mueller_NatureMat09,radu_PRL09,kim_APL09}. In
view of angular momentum conservation a change in the
magnetization $M$ requires transfer of angular momentum from $M$
to some other reservoir. The crystal lattice is a prominent
candidate here, which turns spin-lattice coupling into an
essential, but barely investigated interaction in ultrafast
magnetization dynamics.

In this letter we report on laser-induced magnetization dynamics
in the lanthanide ferromagnets Gd and Tb. By time-resolved x-ray
magnetic circular dichroism (XMCD) at the M$_5$ absorption edges
we probe directly the $4f$ magnetic moment, out of reach for
magneto-optical techniques. We identify for both materials two
separate demagnetization processes, a slower quasi-equilibrium one
and an ultrafast one active in the electronically excited state.
The time constants for the slower process differ for the strong
direct spin-lattice coupling in Tb ($8$~ps) and the weaker
indirect interaction in Gd ($40$~ps). The ultrafast process agrees
for both elements ($0.74$ vs. $0.76$~ps) and is active while hot
electrons are present. It involves an enhancement of the indirect
spin-lattice coupling, which leads to a pronounced increase in the
momentum transfer rates from the magnetization to the lattice in
Gd by as much as $50$ times.

\begin{figure} \centering
 \includegraphics[width=0.95\columnwidth]{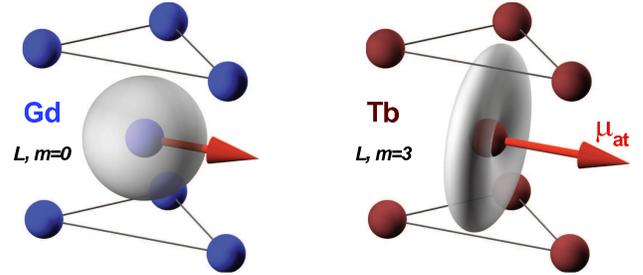}
\caption{\label{Fig1} (color online) Orbital wave-function
distributions within an hcp unit cell for $L, m=0$ and $3$. The
$m=3$ non-spherical distribution of Tb couples to the ion cores
via single ion anisotropy, which is absent for the spherical $m=0$
state of Gd. }
\end{figure}
The heavy lanthanides Gd ($4f^7$) and Tb ($4f^8$) are well known
for their magnetic properties as a function of occupation of the
$4f$ orbital. While the spin quantum number $S$ decreases as the
$4f$ shell is more than half filled (Gd $S=7/2$, Tb $6/2$), the
orbital quantum number $L$ increases (Gd~$L=0$, Tb~$3$)
\cite{coqblin77}. The magnetic moment per atom $\mu_{at}$ follows
Hund's rules (Gd $7.55~\mu_B$, Tb $9.34~\mu_B$
\cite{koehler_JAP65}), where the excess from the integer value is
attributed to spin polarization of the $5d6s$ valence electrons.
Fig.~\ref{Fig1} depicts the $L,m=0$ and $L,m=3$ angular
distribution of the $4f$ orbital of Gd and Tb, respectively; $m$
is the magnetic quantum number. A pronounced coupling of the
orientation of $\mu_{at}$ to the neighbouring ion cores and hence
to the lattice follows for Tb from the non-spherical $4f$
distribution since spin-orbit interaction couples the direction of
the spin moment to the $4f$ orbital. Such a non-spherical
distribution links a rocking of the atomic magnetic moment
directly to a lattice vibration and vice versa. For the spherical
distribution of the half filled Gd $4f$ shell this direct coupling
is absent (Fig.~\ref{Fig1}). Indeed, the magnetic anisotropy
constant $K_2$ describing the energy required to rotate $M$ with
respect to the basal plane of the hcp lattice is in Gd more than
two orders of magnitude smaller than in Tb \cite{coqblin77}. Also
magnon excitations reflect this difference in $L$. Avoided
crossings in the magnon dispersion of Tb explained by
magnon-phonon coupling \cite{jensen_71} are absent in Gd
\cite{koehler_PRL70,melnikov_JPhysD08}. The magnetic anisotropy in
Gd is, however, non-zero due to $4f-5d$ coupling and the
spin-orbit interaction of $5d$ electrons \cite{colarieti_PRL03}.
We refer to such a valence electron mediated spin-lattice coupling
as indirect.

\begin{figure} \centering
\includegraphics[width=0.8\columnwidth]{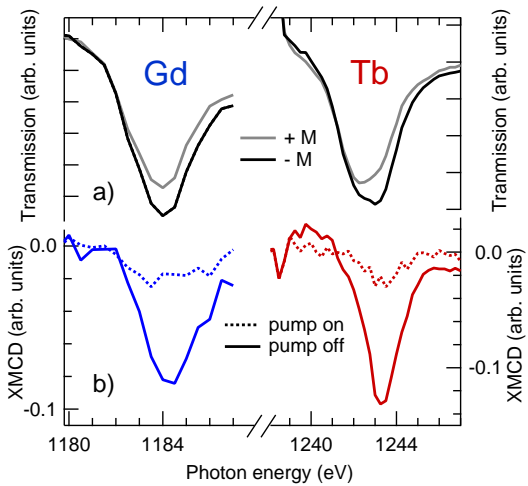}
\caption{\label{Fig2} (color online) a) X-ray transmission at the
M$_5$ absorption edges of Gd and Tb films recorded for opposite
magnetization direction (black and gray lines) with $10$~ps
circularly polarized x-ray pulses. b) XMCD signals of Gd and Tb
before and $200$~ps after laser excitation (solid and dotted
lines).}
\end{figure}

Optical pump -- x-ray probe experiments were performed at the
femtosecond slicing facility of BESSY II \cite{stamm_NatMat08}.
The $5d6s$ valence electrons were excited by 1.5~eV laser pulses
of 50~fs duration at a fluence of $F=3-5$~mJ/cm$^2$ with the
sample held in an applied magnetic field of $5$~kOe at an
equilibrium temperature of $140$~K. We measured x-ray transmission
for poly-crystalline Y(50~nm)/R(10~nm)/Y(5~nm) films grown on a
free-standing $0.5~\mu$m thick Al substrate; R = Gd,Tb. The x-ray
photon energy was tuned to resonantly excite the $3d_{5/2}$
core-level electrons to the unoccupied $4f^{\downarrow}$ states
with a binding energy of $4$~eV above $E_{\mathrm{F}}$
\cite{erskine_73}. Since optical transitions between $4f$ and $5d$
require photon energies far above 1.5~eV, $4f$ levels do not
participate in the optical excitation \cite{erskine_76} and can be
used as a reliable monitor of $M$ \cite{starke_XMOKE_PRL01};
pump-induced refilling of $4f$ levels and saturation effects do
not affect our measurements.

Figure~\ref{Fig2}a shows the transmission spectra for $M$ parallel
($+$) and antiparallel ($-$) to the helicity of circularly
polarized x-ray pulses without laser excitation. XMCD is
determined from the difference of the absorption for opposite $M$.
Comparing XMCD signals before and $200$~ps after laser excitation
(Fig.~\ref{Fig2}b) exhibits a pronounced pump-induced change. The
sum of the spectra (not shown) remains unaffected even though the
temperature is increased by the optical excitation. This
guarantees that the change in XMCD is a purely magnetic effect.

\begin{figure}
\includegraphics[width=.8\columnwidth]{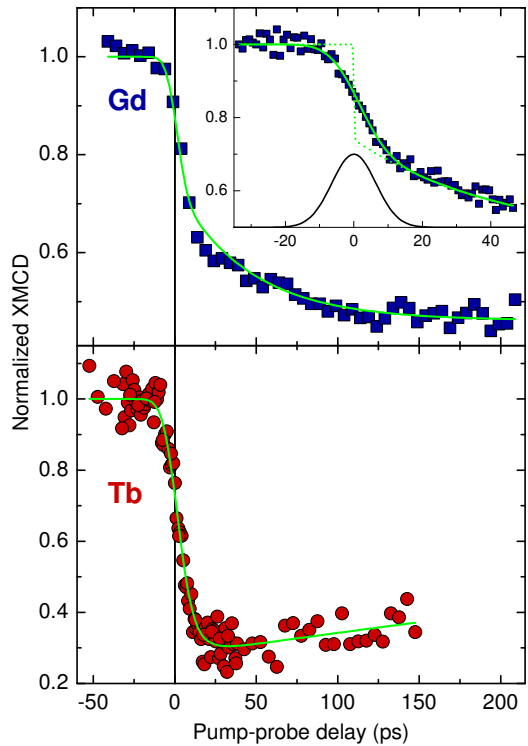}
\caption{\label{Fig3} (color online) Time-dependent XMCD signals
for Gd (top) and Tb (bottom) measured by $10$~ps x-ray probe and
$50$~fs laser pump pulses. Solid lines indicate fits to the data.
The inset depicts Gd data in a smaller time window with the actual
time-resolution of $16$~ps indicated. The biexponential fit (solid
line) highlights the two step demagnetization process and the
dashed line indicates the behavior expected for an instantaneous
first step.}
\end{figure}

We proceed to the magnetization dynamics and analyze the
time-dependence of the XMCD signal. At first, we employed x-ray
pulses of about 10~ps duration, which are available in the
low-$\alpha$ operation mode of BESSY~II \cite{abobakr_PRL02}.
Fig.~\ref{Fig3} depicts the time-dependent XMCD signal for Gd and
Tb  normalized to the value before optical excitation. For both
materials we find a pronounced demagnetization, but the detailed
behavior is different. For Gd the minimum of $M$ is reached after
200~ps in a two-step process. The inset indicates that about half
of the final demagnetization occurs within the 10~ps pulse
duration of the x-ray pulse, while the second process lowers $M$
until 200~ps. We fit the Gd data by a biexponential decay
convoluted with the x-ray pulse duration and determine a
characteristic time constant of
$\tau_{\mathrm{eq}}^{\mathrm{Gd}}=40\pm10$~ps for the slower
process. Electrons and phonons have equilibrated after 1~ps
\cite{bovensiepen_JPCM07}. Therefore, we refer to delays $>1$~ps
as a quasi-equilibrium and $<1$~ps as an electronically excited
state. The obtained $\tau_{\mathrm{eq}}^{\mathrm{Gd}}$ is
characteristic for the weak indirect spin-lattice coupling in Gd
($L=0$, c.f. Fig.~\ref{Fig1}) in quasi-equilibrium, since it is
$\gg$1~ps. Our finding substantiates previous experimental and
theoretical results
\cite{vaterlaus_PRL91,melnikov_PRL08,Huebner96}. From the change
in $M$ stemming from this quasi-equilibrium process at a delay of
$\tau_{\mathrm{eq}}^{\mathrm{Gd}}$ we determine an angular
momentum transfer rate of
$\sigma_{\mathrm{eq}}^{\mathrm{Gd}}=0.026_{-0.005}^{+0.009}~
\mu_{\mathrm{B}}/\mathrm{ps}$ considering $M$ at 140
K~\cite{drulis_91}. In Tb the minimum of $M$ is reached already
after 20~ps indicating a faster demagnetization, which is a
consequence of the direct spin-lattice coupling ($L=3$, c.f.
Fig.~\ref{Fig1}). The cooling mediated recovery of the initial
magnetization is described by an exponential behavior during
several $100$~ps. In Gd diffusive cooling and  slow
demagnetization occur on similar time scales and lead to a
plateau; in Tb cooling occurs after demagnetization and a recovery
of $M$ is observed at delays $>20$~ps.

\begin{figure}
\includegraphics[width=.8\columnwidth]{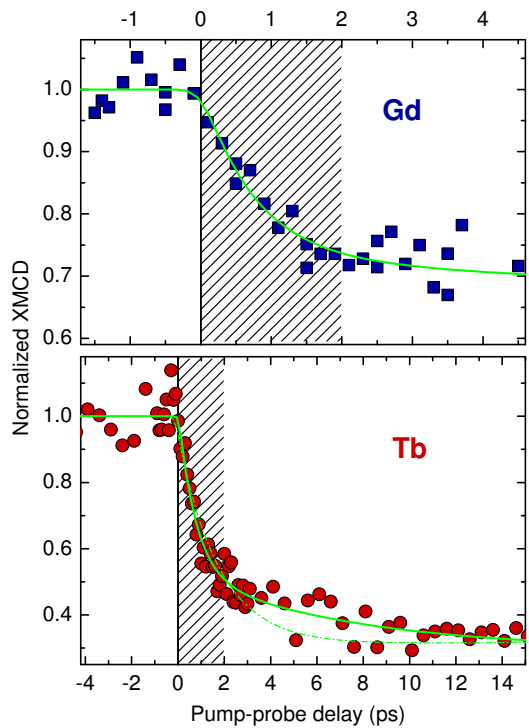}
\caption{\label{Fig4} (color online) Time-dependent XMCD signals
for Gd (top) and Tb (bottom) measured with fs x-ray pulses. Note
the different time intervals. Solid lines depict biexponential
fits determined by simultaneously fitting the fs and ps
time-resolved data (cf. solid lines in Fig.~\ref{Fig3}). A single
exponential (dash-dotted line) for Tb yields an unsatisfactory
fit.}
\end{figure}

Now three questions remain open. (i) What is the fast
demagnetization time scale in Gd? (ii) Does Tb also show two
distinct demagnetization time scales and if yes (iii) do both
differ with respect to Gd? To answer these questions we employed
fs x-ray pulses which we obtain by femtosecond slicing of the
electron bunches in the storage ring \cite{schoenlein_Science00,
stamm_NatMat08}. Fig.~\ref{Fig4} confirms a clear reduction of $M$
for both elements. In Gd we find after $3$~ps a normalized XMCD
signal of 0.7, identical to the level at which the slower
demagnetization process sets in (inset in Fig.~\ref{Fig3}).
Employing the fs x-ray pulses we resolve the initial, fast
demagnetization process in Gd. Also for Tb we find a sizeable drop
of $M$ within $2$~ps (dashed areas in Fig.~\ref{Fig4}).

To determine the characteristic time scales, the ps  and fs
time-resolved data have been fitted simultaneously by
biexponential functions taking into account the different x-ray
pulse durations (solid lines in Figs. \ref{Fig3} and \ref{Fig4}).
For Tb we obtain $\tau^{\mathrm{Tb}}_{\mathrm{eq}}=8\pm3$~ps,
which translates to an angular momentum transfer rate of
$\sigma_{\mathrm{eq}}^{\mathrm{Tb}}=0.29_{-0.08}^{+0.17}~
\mu_{\mathrm{B}}/\mathrm{ps}$. We explain this process as being
mediated by direct spin-lattice coupling under quasi-equilibrium
conditions persisting at corresponding delays $>1$~ps. This is
much faster than $\sigma_{\mathrm{eq}}^{\mathrm{Gd}}=0.026~
\mu_{\mathrm{B}}/\mathrm{ps}$ determined for the indirect
interaction in Gd, which demonstrates that the direct spin-lattice
coupling stemming from the non-spherical $4f$ orbital distribution
accelerates the demagnetization process in Tb.

Next, we focus on the ultrafast demagnetization process. From our
fits we determine within error bars identical times
$\tau^{\mathrm{Gd}}_{\mathrm{ex}}=0.76\pm0.25~$ps and
$\tau^{\mathrm{Tb}}_{\mathrm{ex}}=0.74\pm0.25~$ps. These times are
shorter than reported for Gd/Fe multilayers \cite{bartelt_APL2007}
and similar to reports on TbFe alloys \cite{kim_APL09}. Since they
are clearly longer than the pulse durations we rule out coherent
processes promoted in Ref.~\cite{bigot_NatPhys09}. Note that our
observations are likewise not compatible with demagnetization via
superdiffusive spin transport \cite{battiato_PRL10}. The
ultrafast, component of the demagnetization is $50$~\% of the
total loss in $M$ and thus much too large to be explained by
transport of the $5d$ valence electrons. We can furthermore
exclude a mere transfer of the magnetic moment from $4f$ to $5d$
electrons because (i) the transferred moment is considerably
larger than the valence electron spin polarization in both
lanthanides. (ii) The $4f-5d$ states are coupled by intra-atomic
exchange \cite{ahuja_PRB94}. Depending on the coupling strength a
transfer between $5d$ and $4f$ magnetic moments or a concomitant
change seems reasonable. We have performed time-resolved
magneto-optical experiments, which probe primarily the valence
band spin polarization. We find a reduction in the transient $5d$
magnetic signal concomitant with the XMCD signal measuring the
$4f$ magnetic moment. Considering that the optically excited
electrons in Gd equilibrate with the crystal lattice during 1~ps
\cite{bovensiepen_JPCM07} the ultrafast demagnetization occurs as
long as the system remains electronically excited. Our results
demonstrate an efficient hot electron mediated momentum transfer
to the lattice and determine the corresponding transfer rates of
magnetic moment to the lattice to
$\sigma_{\mathrm{ex}}^{\mathrm{Tb}}=3.1_{-0.8}^{+1.5}~
\mu_{\mathrm{B}}/\mathrm{ps}$ and
$\sigma_{\mathrm{ex}}^{\mathrm{Gd}}=1.5_{-0.4}^{+0.7}~
\mu_{\mathrm{B}}/\mathrm{ps}$. Compared to the quasi-equilibrium
processes discussed above these rates are 10 and 50 times larger
for Tb and Gd, respectively. We propose that the intraatomic
$4f-5d$ exchange interaction, about 100~meV \cite{ahuja_PRB94},
mediates this acceleration in the electronically excited state, as
by means of the $4f-5d$ coupling a spin-flip scattering process in
the conduction band affects the $4f$ electrons as well and thereby
drives the ultrafast demagnetization via indirect spin-lattice
coupling \cite{koopmans_NatMat10}.

Interestingly, in both $4f$ elements the ultrafast process lasts
longer than in the $3d$ transition metal ferromagnets
\cite{beaurepaire_PRL96, stamm_NatMat08, koopmans_NatMat10},
although the momentum transfer rates are comparable, e.g.
$\sigma_{\mathrm{ex}}^{\mathrm{Ni}}=2.7~
\mu_{\mathrm{B}}/\mathrm{ps}$ for Ni \cite{stamm_NatMat08}. This
originates from the fact that the dominant part of $M$ is
generated by the $4f$ electrons and is considerably larger for Gd
and Tb than for Fe, Co, and Ni. As the magnetic moment carried by
the valence electrons is considerably smaller than the $4f$ one,
several spin flips in the conduction band of the lanthanide
elements are required to obtain the same relative demagnetization
as in a $3d$ ferromagnet, where the magnetic moment resides
completely in the conduction band and is much smaller.

Finally, we compare our results with the model calculations of
Ref.~\cite{koopmans_NatMat10}. In agreement we find laser-induced
demagnetization of lanthanides on two time scales. However, the
orbital momentum of the $4f$ shell cannot be neglected because we
observe
$\sigma^{\mathrm{Tb}}_{\mathrm{eq}}=11\cdot\sigma^{\mathrm{Gd}}_{\mathrm{eq}}$.
Ref.~\cite{koopmans_NatMat10} does not consider direct
spin-lattice coupling, shown here to be essential, and predicts a
figure of merit for the demagnetization time that is proportional
to the ratio of Curie temperature and magnetic moment
$T_{\mathrm{C}}/\mu_{\mathrm{at}}$. Applying this to Gd
($T_{\mathrm{C}}=293$~K,
$\mu_{\mathrm{at}}=7.55~\mu_{\mathrm{B}}$) and Tb
($T_{\mathrm{C}}=225$~K,
$\mu_{\mathrm{at}}=9.34~\mu_{\mathrm{B}}$) suggests that
demagnetization in Gd is faster than in Tb by a factor of 1.6.
Even within our conservative error bars our results cannot support
this estimation; the faster demagnetization times coincide for Tb
and Gd. Furthermore these times compare well with the time scale
of electron-phonon equilibration \cite{bovensiepen_JPCM07}. We
consider therefore that the time interval during which the faster
demagnetization process is active is determined by electron-phonon
interaction. A similar $\tau_{\mathrm{ex}}$ for Gd and Tb is very
plausible since their valence electron and crystal lattice are
widely comparable.

In conclusion magnetization dynamics of the $4f$ moments in Gd and
Tb occurs on two timescales. The slow picosecond timescale is
determined by the equilibrium spin-lattice coupling following the
$4f$ occupation. The fast femtosecond timescale is comparable for
Gd and Tb and shows a pronounced enhancement of the
valence-electron mediated indirect spin-lattice coupling. We
expect this mechanism to be operative also in the $3d$
ferromagnets but hard to unravel due to the delocalized character
of the magnetic moment.

We thank T. Quast, K. Holldack, and R. Mitzner for experimental
support and M. F\"ahnle, A. I. Lichtenstein, T. O. Wehling, and M.
I. Katsnelson for fruitful discussions. Financial support by the
Deutsche Forschungsgemeinschaft, the European Union, the HEC-DAAD,
and the US Dept. of Energy under contract DE-AC02-76SF00515 is
gratefully acknowledged.



\end{document}